\documentclass[authoryear]{elsarticle}
\usepackage{subfig} 
\usepackage{amsmath}
\usepackage{amsfonts}
\usepackage{algorithm}
\usepackage{algpseudocode}

\begin{document}

\title{An external field prior for the hidden Potts model\\with application to cone-beam computed tomography}

\author[sms,ihbi]{Matthew T. Moores\corref{cor1}}
\ead{m.moores@student.qut.edu.au}

\author[ihbi,romc,scs]{Catriona E. Hargrave}
\ead{c.hargrave@qut.edu.au}

\author[ihbi,scs]{Fiona Harden}
\ead{fiona.harden@qut.edu.au}

\author[sms,ihbi]{Kerrie Mengersen\corref{cor2}}
\ead{k.mengersen@qut.edu.au}

\cortext[cor1]{Corresponding author}
\cortext[cor2]{Principal corresponding author}
\address[sms]{Mathematical Sciences School, Queensland University of Technology, P.O. Box 2434, Brisbane, Queensland, 4001, Australia}
\address[ihbi]{Institute for Health and Biomedical Innovation, Queensland University of Technology, Kelvin Grove, Queensland, 4059, Australia}
\address[romc]{Radiation Oncology Mater Centre, Queensland Health, Raymond Tce, South Brisbane, Queensland, 4101, Australia}
\address[scs]{School of Clinical Sciences, Queensland University of Technology, P.O. Box 2434, Brisbane, Queensland, 4001, Australia}

\begin{abstract}
In images with low contrast-to-noise ratio (CNR), the information gain from the observed pixel values can be insufficient to distinguish foreground objects.  A Bayesian approach to this problem is to incorporate prior information about the objects into a statistical model. This paper introduces a method for representing spatial prior information as an external field in a hidden Potts model of the image lattice. The prior distribution of the latent pixel labels is a mixture of Gaussian fields, centred on the positions of the objects at a previous point in time. This model is particularly applicable in longitudinal imaging studies, where the manual segmentation of one image can be used as a prior for automatic segmentation of subsequent images. The model is demonstrated by application to cone-beam computed tomography (CT), an imaging modality that exhibits distortions in pixel values due to X-ray scatter. The external field prior results in a substantial improvement in segmentation accuracy, reducing the mean pixel misclassification rate on our test images from 87\% to 6\%.
\end{abstract}

\begin{keyword}
Bayesian image analysis \sep Hidden Markov random field \sep Ising/Potts model \sep Longitudinal imaging \sep Path sampling
\end{keyword}

\maketitle

\section{Introduction}
\label{s:intro}
Longitudinal imaging is a popular method in a wide range of scientific fields including environment, economics, agriculture, biology, and medicine. Remote sensing is used for long-term monitoring of land use \citep{Strickland2011}, water quality \citep{McClain2009} and economic growth \citep{Henderson2011}. X-ray computed tomography (CT), magnetic resonance imaging (MRI) and ultrasound are used for {\em in vivo} studies of livestock production \citep{Alston2007}, tumour progression \citep{Albanese2013} and neurodegenerative disease \citep{Thompson2004}. These images can exhibit artefacts and distortions such as cloud cover in remote sensing, magnetic field inhomogeneities in MRI and X-ray scatter in CT. The resulting poor contrast-to-noise ratio (CNR) creates difficulties in interpreting the images. Multiple images acquired of the same subject increase the information available compared to that obtained from a single acquisition. In order to perform well in this setting, an image processing algorithm must efficiently incorporate all of the available knowledge that characterises the specific types of images encountered.

A Bayesian statistical model can combine information from multiple images as well as from other sources, such as published studies, previous experiments and expert opinion. Bayesian methods for image analysis were reviewed by \citet{Hurn2003b,Hurn2003a}. \citet{Woolrich2012} has provided a more recent overview of methods for analysing medical images, particularly functional MRI. There have been some interesting developments in the area of informative priors. \citet{Grenander2007} defined an information-theoretical framework for representation of complex anatomical structures and diffeomorphisms. \citet{Shyr2011} used a library of segmented images as an informative prior for a nonparametric Bayesian model.
The prior distribution of the model parameters can be updated dynamically, using the posterior distribution of one image as the prior for the next. For example, \citet{Alston2007} derived informative priors for the means and variances of a hidden Potts model from an adjacent slice of a CT scan, or from an earlier scan of the same subject.

In this paper we demonstrate how sources of spatial information can be incorporated into an external field prior for the hidden Potts model. This prior is derived from a previous image of the same subject, combined with published studies of geometric variation. It can be updated sequentially as each image is processed to build a subject-specific spatial model. The prior can be computed offline and is amenable to massively parallel implementation. We show that it improves the accuracy of image segmentation in the presence of noise.

The remainder of this article is organised as follows. Section \ref{s:model} describes the hidden Potts model and introduces the external field prior. Bayesian computation using Markov chain Monte Carlo (MCMC) is described in Section \ref{s:inference}. We review chequerboard updating, path sampling and the algorithm of \citet{Swendsen1987} in the context of our method. In Section \ref{s:cbct} we illustrate our methodology using cone-beam CT scans of an electron density (ED) phantom. The external field prior is well suited to this application, but it would also be broadly applicable to a range of longitudinal imaging datasets. The article concludes with a discussion.

\section{Hidden Potts model}
\label{s:model}
Digital raster images are discretised into a regular lattice of pixels (or voxels, in 3D). Image segmentation can be viewed as the task of labelling each pixel to identify which parts of the image correspond to objects and which comprise the background. If these objects are contiguous then neighbouring pixels are more likely to share the same label. The \citet{Potts1952} model represents this spatial dependence as a Markov random field (MRF), which is defined in terms of its conditional probabilities:
  \begin{equation}
  \label{eq:Potts}
  \pi(z_i|z_{i \sim \ell},\beta) = \frac{\exp\left\{\beta\sum_{i \sim \ell}\delta(z_i,z_\ell)\right\}}{\sum_{j=1}^k \exp\left\{\beta\sum_{i \sim \ell}\delta(j,z_\ell)\right\}}
  \end{equation}
  where $z_i \in 1 \dots k$ is the label of pixel $i$, $z_{i \sim \ell}$ are the labels of the neighbouring pixels, $\beta$ is the inverse temperature, and $\delta(u,v)$ is the Kronecker delta function. Thus, $\sum_{i \sim \ell}\delta(z_i,z_\ell)$ is a count of the neighbours of pixel $i$ that share the same label.

If the coordinates of $z_i$ are $(x,y)$ in 2 dimensions, then the first-order neighbours are $(x-1,y),(x+1,y),(x,y-1),(x,y+1)$. Pixels situated at the edge of the image domain have less than four neighbors. Likewise, voxels in a regular 3D lattice have a maximum of 6 first-order neighbours. Since diagonal voxels are conditionally independent, the entire image can be partitioned into two blocks such that all of the voxels in block$_1$ are independent given the values of block$_2$ and vice versa \citep[chap. 8]{Winkler2003}. Updating each block separately reduces the dependence in the posterior samples, providing a computational advantage of the first-order neighbourhood restriction. This is discussed further in Section~\ref{s:labels}.

  The inverse temperature of the Potts model governs spatial cohesion. When $\beta = 0$ the pixel labels are independent, while if $\beta > 0$ then adjacent labels are more likely to have the same value. At small values of $\beta$ this has the effect of smoothing the labels and emphasising patterns in the external field. As $\beta \to \infty$ all of the labels have the same value almost surely. Thus, the inverse temperature can have a substantial influence over the labels that are assigned to each pixel.

  A key question when fitting this type of model is how much smoothing is appropriate for the observed data. For this reason we would like to treat $\beta$ as a free parameter and perform posterior inference to draw samples from its distribution. The full conditional distribution of $\beta$ is as follows:
  \begin{equation}
  \label{eq:beta}
  p(\beta|\mathbf{z}) = \mathcal{C}(\beta)^{-1} \pi(\beta) \exp\left\{ \beta \sum_{i \sim \ell \in \mathcal{E}} \delta(z_i,z_\ell) \right\}
  \end{equation}
where $\mathcal{C}(\beta)$ is an intractable normalising constant, $\pi(\beta)$ is the prior for $\beta$, and $\mathcal{E}$ is the set of all unique pairs of neighbours in the lattice. A variety of computational methods have been developed for dealing with the intractability of this distribution, as explained in Section~\ref{s:beta}.

Under the assumption of additive white noise, the pixels that comprise each object have a mean intensity value $\mu_j$ and variance $\sigma^2_j$. The hidden Potts model can thus be viewed as a spatial generalisation of an independent mixture of Gaussians. The latent labels are related to their corresponding pixel values by the observation equation:
  \begin{equation}
  \label{eq:obs}
  \exp\{\alpha_{i,j}\} \;=\; p(y_i|\mu_j, \sigma_j^2, z_i\!=\!j) \;=\; \phi(y_i|\mu_j,\sigma_j^2)
  \end{equation}
where $\phi(y|\mu,\sigma^2)$ is the Normal density function and $\boldsymbol\alpha$ is an external field. This log-likelihood is combined with the spatial prior to determine the full conditional distribution of the pixel labels:
  \begin{equation}
  \label{eq:post_Potts}
p(z_i | z_{i \sim \ell}, \beta, \boldsymbol\mu, \boldsymbol{\sigma^2}, y_i) = \frac{\exp\left\{\alpha_{i,z_i} + \beta\sum_{i \sim \ell}\delta(z_i,z_\ell)\right\}}{\sum_{j=1}^k \exp\left\{\alpha_{i,j} + \beta\sum_{i \sim \ell}\delta(j,z_\ell)\right\}}
  \end{equation}

\subsection{External field prior}
\label{s:xfield}
\begin{figure*}
\subfloat[$\sigma_\Delta=14.6\textrm{mm}$]{\includegraphics[width=57mm]{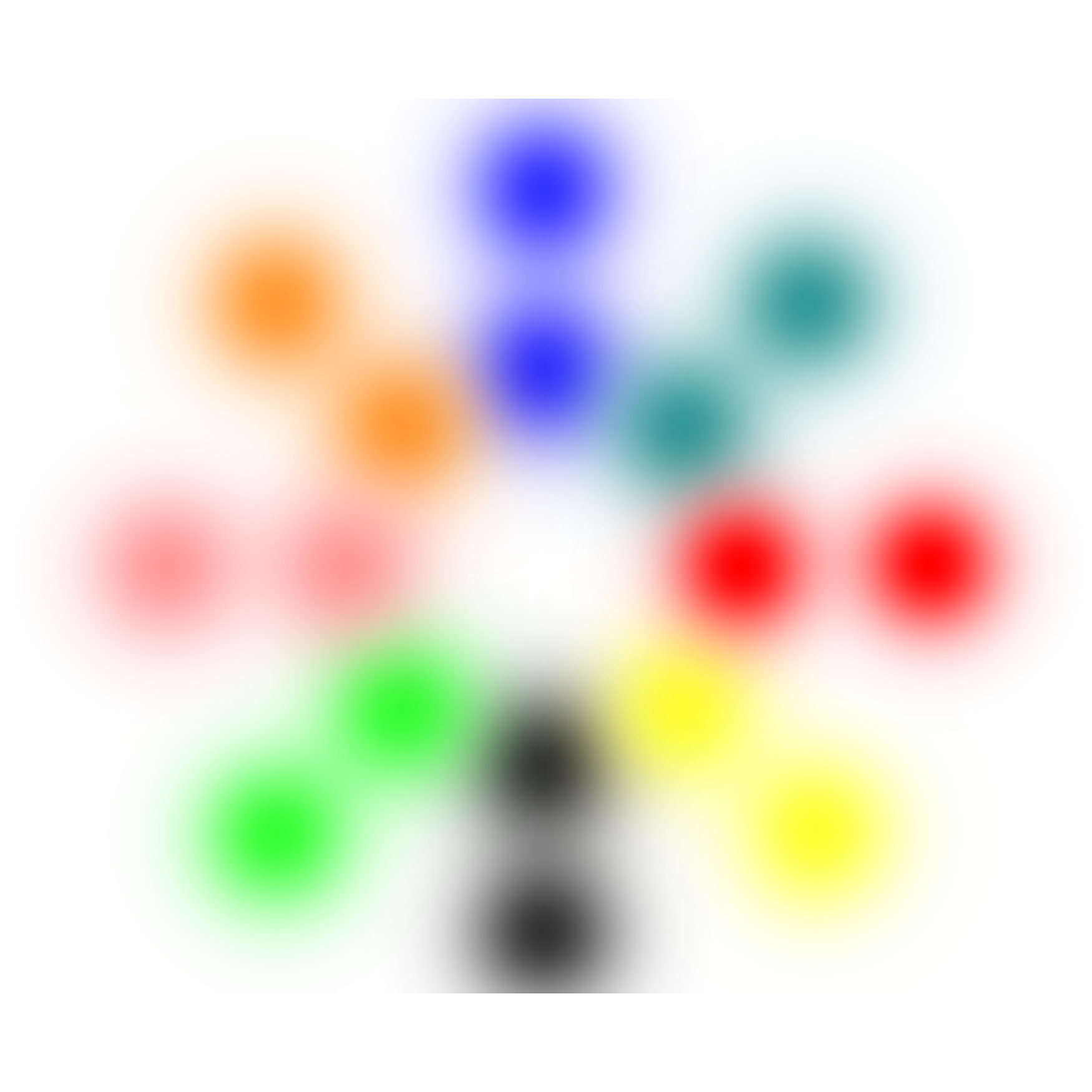}}
\qquad
  \subfloat[$\sigma_\Delta=21.9\textrm{mm}$]{\includegraphics[width=57mm]{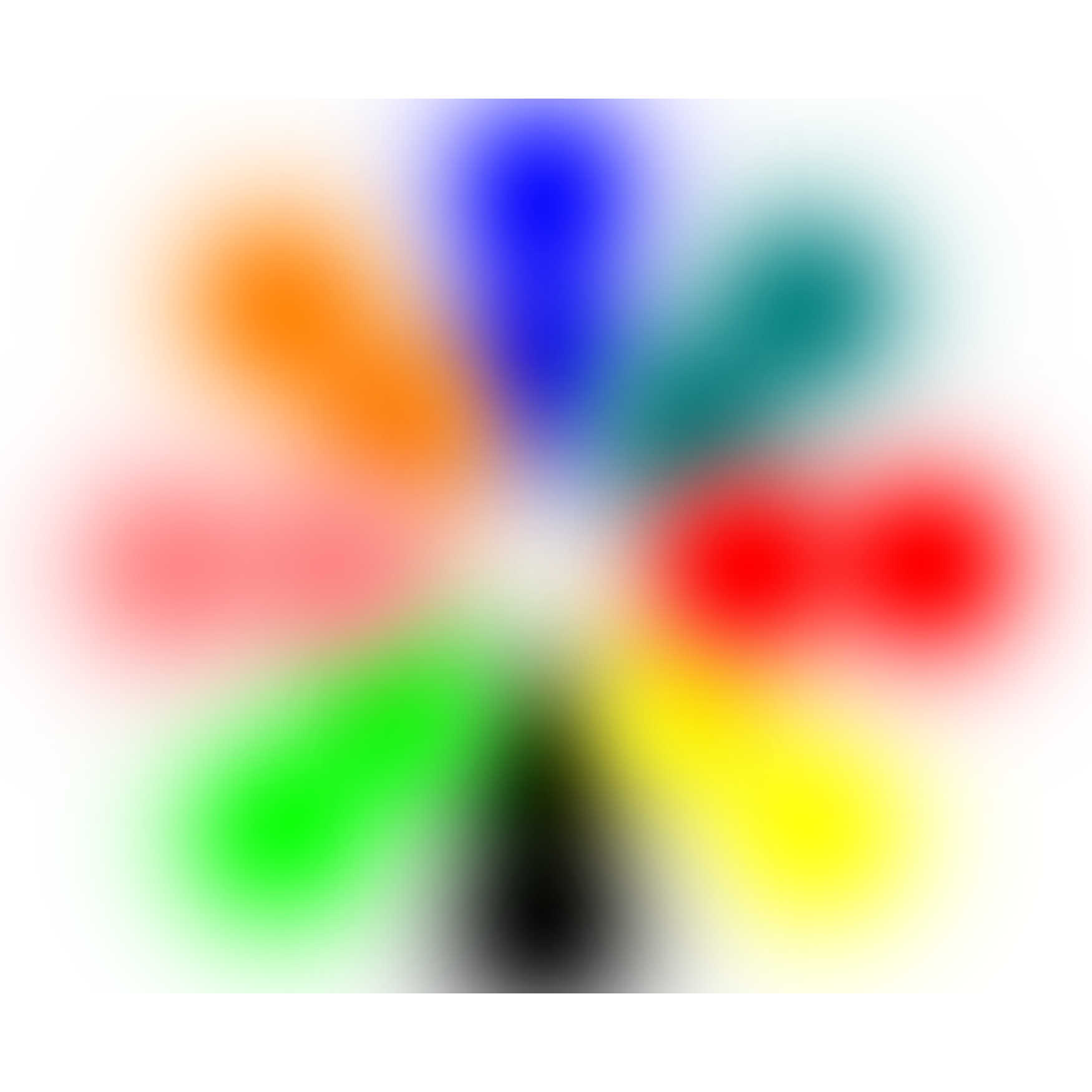}}
\qquad
  \subfloat[$\sigma_\Delta=29.2\textrm{mm}$]{\includegraphics[width=57mm]{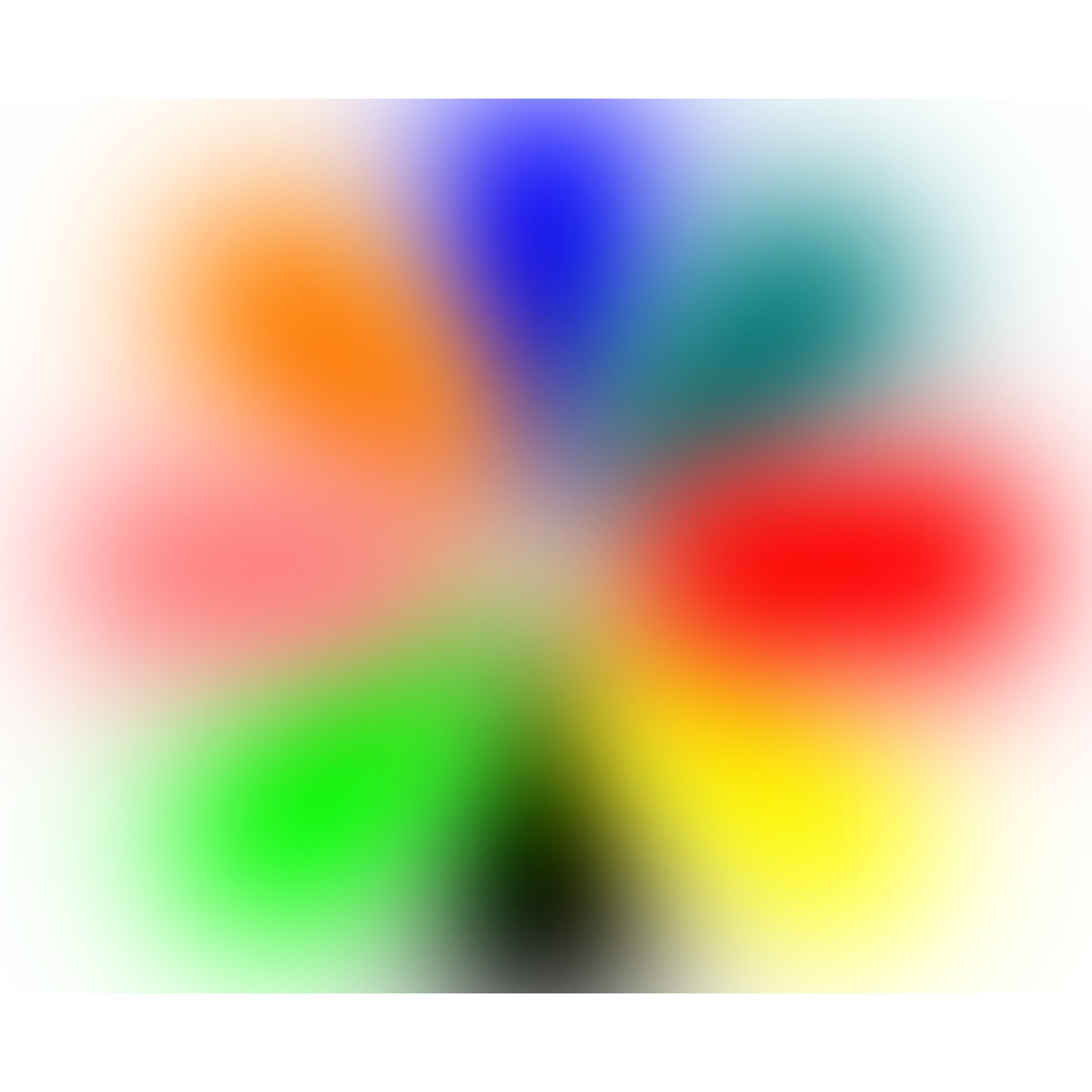}}
\qquad
  \subfloat[$\sigma_\Delta=36.5\textrm{mm}$]{\includegraphics[width=57mm]{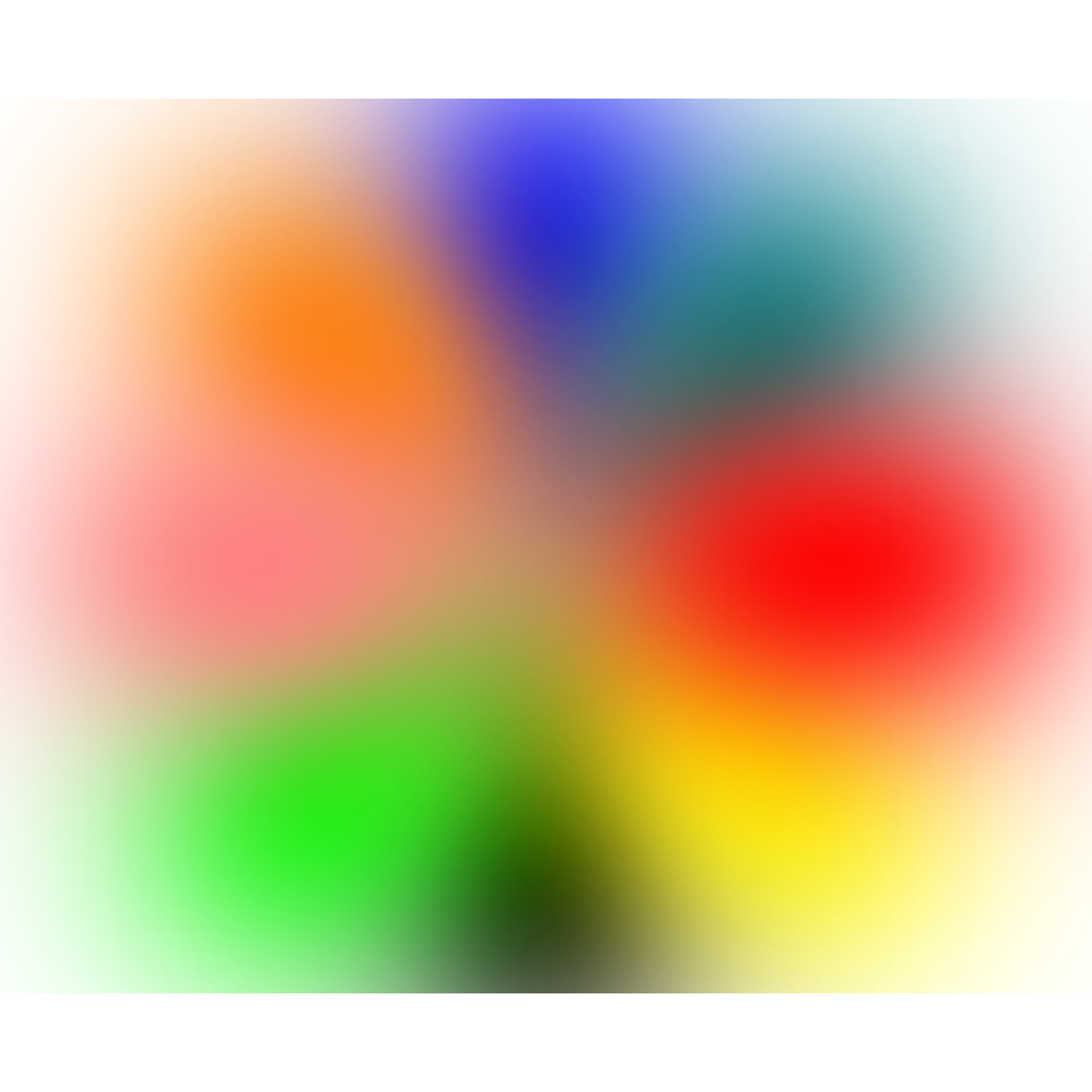}}
% figure caption is below the figure
\caption{External field priors with increasing levels of spatial uncertainty, represented by the standard deviation.}
\label{f:xfield_sigma}
\end{figure*}
In this study, we extend the hidden Potts model by defining a prior $\pi(\boldsymbol\alpha)$ over the external field. This prior incorporates additional information on the unobserved labels. Such information can be derived from a manual segmentation of a previous image, combined with a measure of the spatial uncertainty associated with the labels. Sources of uncertainty include object motion, errors in labelling, and misalignment between images.
  
 For the purpose of illustration, we consider a simple model of isotropic translation in two dimensions. The spatial distribution of pixel labels is represented as a mixture of Gaussians, with one mixture component per pixel. The magnitude of the displacement vector can be calculated for each pixel and the resulting probabilities can be averaged over all of the pixels in the object:
\begin{equation}
\label{eq:gmmdist}
 \pi(\alpha_{i,j}) = \log\left\{ \frac{1}{n_j} \sum_{h \in j} \phi\left(\Delta(h,i)|\mu_\Delta,\sigma^2_\Delta\right) \right\}
\end{equation}
 where
 $n_j$ is the number of pixels in object $j$,
 $h \in j$ are the pixels in object $j$, and
 $\Delta(u,v)$ is the Euclidean distance between the coordinates of pixel $u$ and pixel $v$.
The locations of the pixels $h \in j$ can be obtained from an exemplar image that has been manually labelled. The mean $\mu_\Delta$ is the average spatial displacement of the object between images, while the variance $\sigma^2_\Delta$ represents the degree of uncertainty. Fig. \ref{f:xfield_sigma} illustrates external field priors with increasing levels of spatial uncertainty for the ED phantom that is described in Section~\ref{s:cbct}.

 Since this prior information is independent of the data, it can be incorporated into the Potts model as an external field. The relationship in Eq.~(\ref{eq:post_Potts}) is modified as follows:
\begin{equation}
\label{eq:prior_xfield}
p(z_i | z_{i \sim \ell}, \beta, \boldsymbol\mu, \boldsymbol{\sigma^2}, y_i) = \frac{\exp\left\{\alpha_{i,z_i} + \pi(\alpha_{i,z_i})\right\}}{\sum_{j=1}^k \exp\left\{\alpha_{i,j} + \pi(\alpha_{i,j})\right\}}  \pi(z_i|z_{i \sim \ell},\beta) 
\end{equation}
 where
 $\exp\{\alpha_{i,j}\}$ is the likelihood in Eq.~(\ref{eq:obs}), $\pi(\alpha_{i,j})$ is the external field prior defined in Eq.~(\ref{eq:gmmdist}), and $\pi(z_i|z_{i \sim \ell},\beta)$ is the MRF defined by Eq.~(\ref{eq:Potts}). Furthermore, if the pixel resolution of the image is known in advance, then the values of $\pi(\alpha_{i,j})$ can be calculated offline. This means that much more realistic and sophisticated dynamic models could be employed, in order to estimate the spatial prior distribution from the labels of an exemplar image.

In cases where the pixel resolution is not known in advance, or if there is mismatch between the pixel coordinates of successive images, the discretisation of the Gaussian prior into individual $\pi(\alpha_{i,j})$ values would need to be performed at the model fitting stage. This step is computationally intensive, but since the $\pi(\alpha_{i,j})$ values are independent from each other the operation is embarassingly parallel. Thus the computation of the external field prior would be well suited to implementation in graphical processing units (GPUs) to meet the performance requirements of practical, real world applications.

\subsection{Sequential Bayesian updating}
\label{s:update}
Unlike the atlas-based methods that are currently employed in medical imaging, this external field prior represents within-subject geometric variation. The prior can be updated as each image is acquired to gradually build a subject-specific profile of motion and variability. A machine learning approach to this problem would be to train the model using a corpus of images from a variety of subjects, thus forming a generic picture of the typical subject. However, this fails to account for large differences between individual subjects, thereby conflating within-subject and between-subject variability. Our method develops a subject-specific distribution using a labelled reference image as a starting point, then updating the prior dynamically as more images of the same subject are processed.

\citet{Alston2007} introduced a method for updating the parameters $\mu_j'$ and $\sigma_j^{2\prime}$ of the additive Gaussian noise, using the posterior distribution from one image as the prior for the next. A similar approach could be used for the inverse temperature. Although there is no natural conjugate prior for $\beta$, a distribution from the exponential family could be used to approximate the posterior. A scaled Logistic distribution could represent a posterior with finite support, or a Gamma could be used for $\beta > 0$. When the external field prior is modelled as a mixture of Gaussian fields, its parameters can also be updated in a similar manner.

The location of the region of interest in subsequent images is uncertain. For this reason, we apply weights to the pixel labels according to their posterior probability. A Monte Carlo estimate of this probability can be obtained by dividing the number of times that a given pixel was allocated a specific label by the total number of MCMC iterations (after burn-in):
\begin{equation}
\label{eq:weights}
w_{i,j} = \frac{1}{n_{iter}} \sum_{t=1}^{n_{iter}} \delta(z_i^{(t)}, j)
\end{equation}
thus $w_{i,j} \ge 0$ $\forall j \in 1\dots k$ and $\sum_{j=1}^k w_{i,j} = 1$ for each pixel $i$. It will usually be necessary to adjust this estimate to account for autocorrelation of the Markov chain. Even with the chequerboard updating scheme described in Section~\ref{s:labels}, mixing of MCMC for the hidden Potts model can be less than ideal.

The sufficient statistics for $\pi\left(\mu_\Delta | \sigma_\Delta^2\right) \sim \mathcal{N}\left(m, \sigma_\Delta^2 / \nu\right)$ and $\pi\left(\sigma_\Delta^2\right) \sim \mathcal{IG}\left(\frac{\nu}{2}, \frac{\nu s^2}{2}\right)$ are the sum of the weights, the weighted mean of the Euclidean distances, and the weighted variance:
\begin{eqnarray}
\hat\nu_{\Delta j} &=& \sum_{i=1}^n w_{i,j} \\
\label{eq:upd_mu} \hat{m}_{\Delta j} &=& \frac{1}{\hat\nu_{\Delta j}} \sum_{i=1}^n w_{i,j} \min_{h \in j} \Delta(h,i) \\
\label{eq:upd_sd} \hat{s}_{\Delta j}^2 &=& \frac{1}{\hat\nu_{\Delta j}} \sum_{i=1}^n w_{i,j} \left(  \min_{h \in j} \Delta(h,i) - \hat{m}_{\Delta j}\right)^2
\end{eqnarray}
We use the minimum Euclidean distance between each pixel $i$ and the reference coordinates $h \in j$ because the individual pixels are exchangeable. Calculating the average distance would require $\mathcal{O}(n_j \times n)$ operations, which is infeasible for nontrivial images. The minimum distances can be precomputed in linear time for a regular lattice using the algorithm of \citet{Maurer2003}. If all of the subsequent images have the same coordinates and pixel resolution, then the Euclidean distance transform only needs to be computed once.

Although this approach is computationally efficient, the use of the minimum distances introduces a bias towards underestimation. An additional factor could be added to Eq.~(\ref{eq:upd_mu}) and (\ref{eq:upd_sd}) to compensate for the bias, but this would also increase the computational cost. For example, $\min_{h \in j} \Delta(h,i)$ could be augmented with the average distance between the pixels inside the object:
\begin{equation}
\frac{1}{n_j^2} \sum_{g \in j} \sum_{h \in j} \Delta(g,h)
\end{equation}
This would be tractable for small objects where $n_j \ll n$.

The hyperparameters of the external field prior can be updated in the usual manner for the Gaussian distribution:
\begin{eqnarray}
\nu_{\Delta j}' &=& n_j^{\circ} + \hat\nu_{\Delta j} \\
m_{\Delta j}' &=& \frac{1}{\nu_{\Delta j}'}\left(n_j^\circ \mu_\Delta^\circ + \hat\nu_{\Delta j} \hat{m}_{\Delta j}\right) \\
s_{\Delta j}^{2\prime} &=& \frac{1}{\nu_{\Delta j}'}\left( n_j^\circ \sigma_\Delta^{2\circ} + \hat\nu_{\Delta j} \hat{s}_{\Delta j}^2 + \frac{n_j^{\circ} \hat\nu_{\Delta j}}{\nu_{\Delta j}'}\left( \hat{m}_{\Delta j} - m_{\Delta j}' \right)^2 \right)
\end{eqnarray}

One option for updating the individual $\pi(\alpha_{i,j})$ values for each pixel would be to take the expectations $\mathbb{E}\left[ \sigma_\Delta^{2\prime} | \mathbf{y} \right]$ and $\mathbb{E}\left[\mu_{\Delta j}' | \mathbf{y} \right]$ and plug them in to Eq.~(\ref{eq:gmmdist}). This would be the simplest approach and the least computationally intensive, but replacing the posterior distributions with point estimates would underestimate the level of uncertainty. Alternatively, the external field prior could be simulated using MCMC, either offline or during model fitting. As previously mentioned, this would be the preferred option if a more complex dynamic model was used.

\section{Bayesian computation}
\label{s:inference}

We use natural conjugate priors for the mean and variance of each mixture component, so updates can be drawn using Gibbs sampling. However, this approach is unsuitable for the other parameters of interest. Instead, we use chequerboard updating for the latent labels and path sampling for the inverse temperature. Our implementation of these algorithms using RcppArmadillo \citep{Eddelbuettel2014} is available as an \textsf{R} source package with the electronic version of this article. The details of these algorithms are described below.

\subsection{Chequerboard updating}
\label{s:labels}
\begin{algorithm}
\begin{algorithmic}[1]
\ForAll{blocks $b$}
\ForAll{pixels $i \in b$}
\ForAll{labels $j \in 1 \dots k$}
\State Compute $\lambda_j \gets p(j|y_i,\beta)$ according to equation~(\ref{eq:prior_xfield})
\EndFor
\State Draw $z_i \sim \textrm{Multinomial}(\lambda_1,\dots,\lambda_k)$
\EndFor
\EndFor
\end{algorithmic}
\caption{Chequerboard updating for $\mathbf{z}$}
\label{alg:blockGibbs}
\end{algorithm}
The posterior distribution of the pixel labels is highly correlated, so single-site Gibbs sampling exhibits very slow mixing \citep{Higdon1998}. For this reason we divide the lattice into blocks, as mentioned in Section~\ref{s:model}. The pixels within each block are independent of each other, so they can be updated in parallel. The auxiliary variable $b_i \in \{0,1\}$ allocates each pixel to a block. \citet{Winkler2003} describes this partially synchronous algorithm as chequerboard updating, since the alternating allocations of pixels to blocks forms a chequerboard pattern in two dimensions. The dissertation of \citet{Feng2008} provides a more detailed description and also generalises this algorithm to 3D, as well as to other neighbourhood schemes.

\subsection{Path sampling}
\label{s:beta}
\begin{figure}
   \includegraphics[width=5.3in]{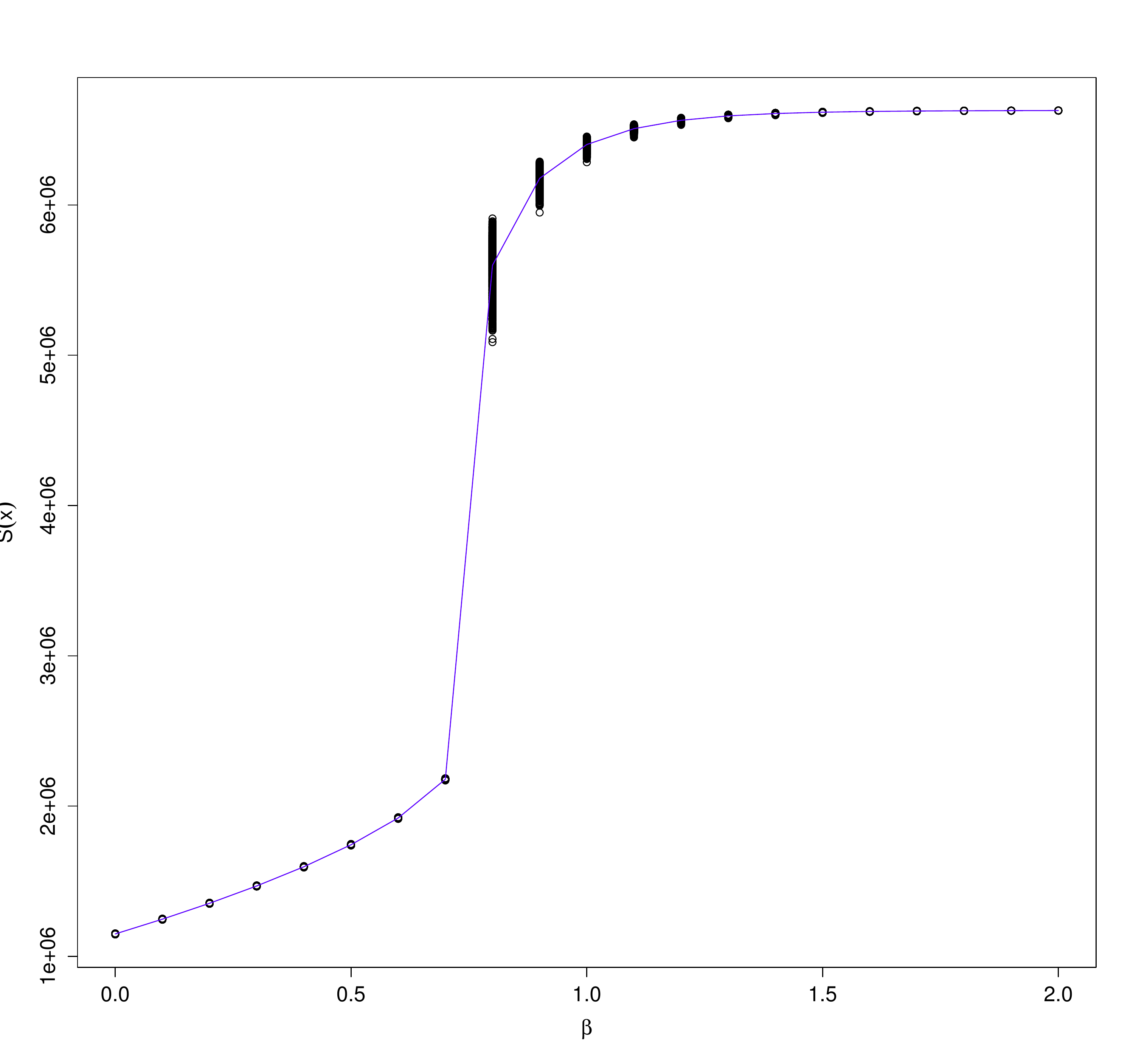}
\caption{Approximation of the sufficient statistic by simulation for fixed values of $\beta$, with linear interpolation.}
\label{f:path}
\end{figure}
The full conditional distribution of the inverse temperature (Eq.~\ref{eq:beta}) involves an intractable normalising constant, so there is no closed-form solution for sampling from it directly. \citet{Gelman1998} derived an approximation to the log ratio of normalising constants using the path sampling identity:
\begin{equation}
\label{eq:path}
\log\left\{\frac{\mathcal{C}(\beta^\circ)}{\mathcal{C}(\beta')}\right\} = \int_{\beta'}^{\beta^\circ} \mathbb{E}\,_{\mathbf{z}|\beta} \left[\mathrm{S}(\mathbf{z})\right] \, \mathrm{d}\beta
\end{equation}
where $\beta^\circ$ is the current value of the parameter,
$\beta'$ is the Metropolis-Hastings (M-H) proposal,
$\mathrm{S}(\mathbf{z}) = \sum_{i \sim \ell \in \mathcal{E}} \delta(z_i,z_\ell)$ is the sufficient statistic of the Potts model,
and $\mathbb{E}\,_{\mathbf{z}|\beta}$ is the expectation with respect to the distribution of $\mathbf{z}$ given $\beta$. The value of this definite integral can be approximated by simulating from the MRF defined by Eq.~(\ref{eq:Potts}) for fixed values of $\beta$ and then interpolating between them. We used the Swendsen-Wang algorithm for simulating from $\mathbf{z}|\beta$, as explained in Section~\ref{s:SW}. Fig.~\ref{f:path} illustrates linear interpolation of $\mathrm{S}(\mathbf{z})$ on a 3D lattice for $k=9$ and $\beta$ ranging from 0 to 2 in increments of 0.1.
\begin{algorithm}
\begin{algorithmic}[1]
\State Draw random walk proposal $\beta' \sim q(\beta'|\beta^\circ)$
\State Estimate $\mathrm{S}(\mathbf{z}|\beta^\circ)$ and $\mathrm{S}(\mathbf{z|\beta'})$ by interpolation
\State Evaluate the definite integral in equation~(\ref{eq:path})
\State Calculate the log M-H acceptance ratio:
  \begin{equation}
  \label{eq:mhRatio}
  \log\{\rho\} = \log\left\{\frac{\mathcal{C}(\beta^\circ)}{\mathcal{C}(\beta')}\right\} + (\beta' - \beta^\circ)\mathrm{S}(\mathbf{z})
  \end{equation}
\State Draw $u \sim \mathrm{Uniform}[0,1]$
\If{$u < \rho$}
\State set $\beta \gets \beta'$ \EndIf
\end{algorithmic}
\caption{Path sampling for $\beta$}
\label{alg:path}
\end{algorithm}

Path sampling is explained in further detail by \citet{Hurn2003b} and \citet[chap. 8]{Marin2007}. This algorithm has an advantage over auxiliary variable methods such as the exchange algorithm \citep{Murray2006} or ABC \citep{Grelaud2009} because the additional simulations are performed prior to fitting the model, rather than at each iteration. This is particularly the case when analysing multiple images that all have approximately the same dimensions, as in the example of Section~\ref{s:cbct}.

\subsection{Swendsen-Wang}
\label{s:SW}
MCMC takes longer to converge for larger values of $\beta$, since the latent labels tend to keep the same value for many iterations. The algorithm of \citet{Swendsen1987} avoids this problem by coalescing adjacent pixels with the same label into clusters, then updating all of the labels within a cluster to the same value.
\begin{algorithm}
\begin{algorithmic}[1]
\ForAll{edges connecting adjacent pixels $i \sim j \in \mathcal{E}$}
\State Compute edge potentials $\lambda_{i \sim j} \gets 1 - \exp\{-\beta \delta(z_i,z_j)\}$
\State Draw bond $b_{i \sim j} \sim \mathrm{Bernoulli}(p = \lambda_{i \sim j})$
\EndFor
\State Coalesce bonds into clusters $c$
\ForAll{clusters}
\State Draw a new label $z'$ uniformly from $1 \dots k$
\ForAll{pixels $i \in c$}
\State $z_i \gets z'$
\EndFor
\EndFor
\end{algorithmic}
\caption{Swendsen-Wang for $\mathrm{S}(\mathbf{z}|\beta)$}
\label{alg:sw}
\end{algorithm}

A disadvantage of Algorithm~\ref{alg:sw} is that it performs poorly in the presence of a strong external field \citep{Hurn1997}. Thus it is unsuitable for posterior inference, but can be used for simulating from $\mathbf{z}|\beta$ as required for path sampling.

\section{Illustration: cone-beam CT}
\label{s:cbct}
There are two essential ingredients for the external field prior: reference coordinates, which we derive from a manually-labelled image; and spatial uncertainty, which we model using published studies of geometric variation. We have chosen to illustrate our method using cone-beam CT scans because both of these elements are readily available in a clinical context. However, it should be noted that this method could also be applied to longitudinal imaging studies in many other scientific fields, such as satellite remote sensing or confocal laser microscopy.

Flat-detector, cone-beam CT was introduced in the last decade for image-guided medical interventions \citep{Jaffray2002,Kalender2011}. Since then, it has been widely adopted for clinical use in dental surgery, brachytherapy and external-beam radiotherapy. Cone-beam CT produces a 3D image of the patient similar to conventional, fan-beam CT.  The advantage of cone-beam CT over fan-beam CT is that the X-ray source and the detector panel are mounted on retractable arms that rotate around the image subject, enabling the image to be obtained {\it in situ} during a medical procedure.

 The disadvantage of cone-beam CT is that it is more susceptible to artefacts induced by X-ray scatter \citep{Siewerdsen2001} or high density objects such as metal implants \citep{Mueller2009} compared to other imaging modalities. The increased magnitude of X-ray scatter in cone-beam CT is due to the larger area that is exposed to the X-ray beam. This leads to shading artefacts that manifest as inhomogeneities in the pixel values. Metal-induced artefacts can be due to implanted gold fiducial markers or surgical clips inside the patient. These manifest as streaking and banding in the image. These distortions result in a lower CNR and thus many methods that can be successfully used for analysing conventional, fan-beam CT or other imaging modalities encounter difficulties when applied to cone-beam CT.

To evaluate the segmentation accuracy of our method, we applied it to 27 cone-beam CT scans of a commercially available electron density (ED) phantom \citep{CIRS2011}. The ED phantom is manufactured from epoxy and contains cylindrical inserts that mimic the X-ray absorption of human tissue: lung (inhale); lung (exhale); adipose (fatty tissue); breast (50\% fat); water-equivalent solid; muscle; liver; spongy (trabecular) bone; and dense (cortical) bone. Since the geometry and density of this object are known, the segmentation can be compared to the ground truth to evaluate accuracy. This provides an advantage over using scans of patients to evaluate the method because in clinical data the true segmentation is unknown \citep{Bouix2007}. A reference segmentation can be provided by a clinical expert, but this also introduces a source of error. Due to the irregularity and complexity of the medical images, there can be significant discrepancies between segmentations of the same image by different experts \citep{Luetgendorf-Caucig2011,Weiss2010}.

 The 27 cone-beam CT scans were taken with the inner ring of imaging phantom inserts rotated by between $0^{\circ}$ and $16^{\circ}$, corresponding to a translation of between $0\mathrm{mm}$ and $25\mathrm{mm}$. These displacements were intended to mimic the physiological variability observed in daily cone-beam CT scans of prostate cancer patients, as studied by \citet{Frank2008}. We chose to use this as the basis of our experimental design because image-guided radiotherapy for prostate cancer is the most common application of cone-beam CT imaging.

 The cone-beam CT scans were acquired using a Varian linear accelerator with On-Board Imager (OBI) according to the same clinical protocol that is used for scanning prostate cancer patients. A half bow-tie filter was used to achieve a 450mm field of view. The dimensions of the reconstructed voxels were $0.88\times0.88\times2\mathrm{mm}$. The images were cropped using external image processing software \citep{Schneider2012} to conform to the dimensions of the phantom: $376\times308$ pixels and $23$ slices, or approximately $330\times270\times46\textrm{mm}$. Summary statistics (observed mean and standard deviation) for each cone-beam CT scan are summarised in the supplementary material accompanying the electronic version of this paper. An axial slice from one of the cone-beam CT scans is shown in Fig.~\ref{f:res_cbct}.

\subsection{Priors}
\label{s:priors}
 The external field prior was precomputed according to Eq.~(\ref{eq:gmmdist}). This prior was centred on the geometry of the ED phantom at 0 offset. We used a mean displacement of $1.2\mathrm{mm}$ with standard deviation $7.3\mathrm{mm}$. This spatial variability corresponds to the prostate motion observed by \citet{Frank2008} in their study of 15 radiotherapy patients. The external field is illustrated in Fig.~\ref{f:res_xfield}.

We also used informative priors for the means and variances of the mixture components, using a similar method to \citet{Alston2007}.  To determine the distribution of pixel intensities for each tissue type, we obtained an independent set of 26 cone-beam CT scans of the ED phantom. The resulting priors for $\mu_j \sim \mathcal{N}(m_j,\varphi_j^2)$ and $\sigma_j^2 \sim \mathrm{IG}\left(\frac{\nu}{2},\frac{\nu s^2}{2}\right)$ are listed in Table~\ref{t:priors}.

\begin{table}
\caption{Informative priors for $\mu_j$ and $\sigma_j^2$ of the pixel intensity}
\label{t:priors}
\begin{tabular}{lrrrr}
\hline\noalign{\smallskip}
Tissue Type & $m_j$ & $\varphi_j^2$ & $\nu_j$ & $s_j^2$  \\
\noalign{\smallskip}\hline\noalign{\smallskip}
Lung (inhale) & $-612.6$ & $26.88^2$ & $25$ & $90.06^2$ \\
Lung (exhale) & $-495.8$ & $26.88^2$ & $25$ & $89.16^2$ \\
Adipose       & $-316.8$ & $26.88^2$ & $25$ & $79.36^2$ \\
Breast        & $-295.9$ & $26.88^2$ & $25$ & $67.42^2$ \\
Water         & $-294.5$ & $26.88^2$ & $25$ & $152.0^2$ \\
Muscle        & $-263.3$ & $26.88^2$ & $25$ & $71.55^2$ \\
Liver         & $-259.6$ & $26.88^2$ & $25$ & $88.50^2$ \\
Spongy Bone   & $-191.1$ & $26.88^2$ & $25$ & $87.36^2$ \\
Dense Bone    &  $77.9$ & $26.88^2$ & $25$ & $89.94^2$ \\
\noalign{\smallskip}\hline
\end{tabular}
\end{table}

\subsection{Results}
\label{s:results}
% For two-column wide figures use
\begin{figure*}
  \subfloat[External field prior]{\label{f:res_xfield} \includegraphics[width=57mm]{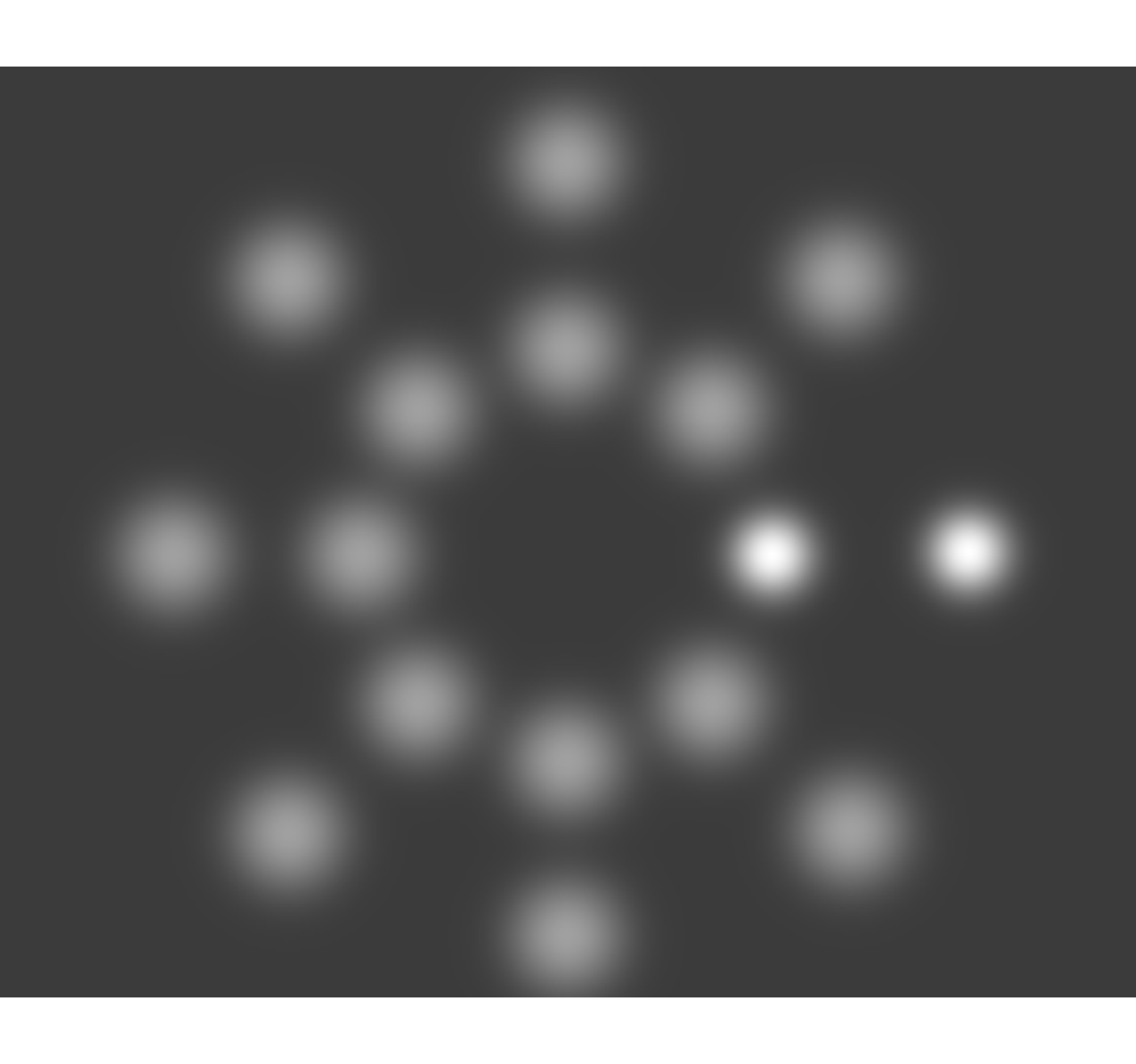}}
\qquad
  \subfloat[Cone-beam CT scan]{\label{f:res_cbct} \includegraphics[width=57mm]{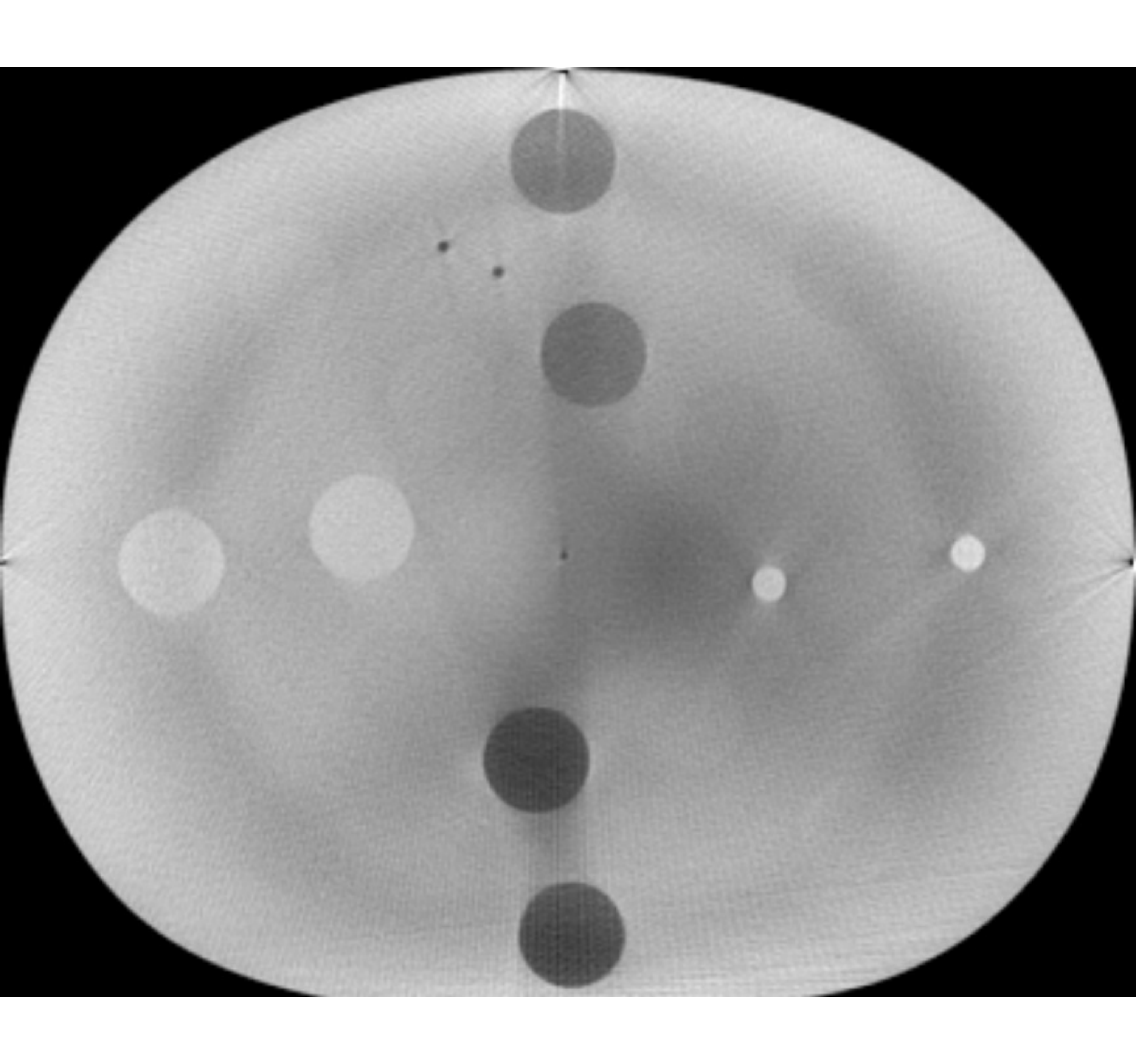}} \\
\qquad
  \subfloat[Segmentation (without external field)]{\label{f:res_noprior} \includegraphics[width=57mm]{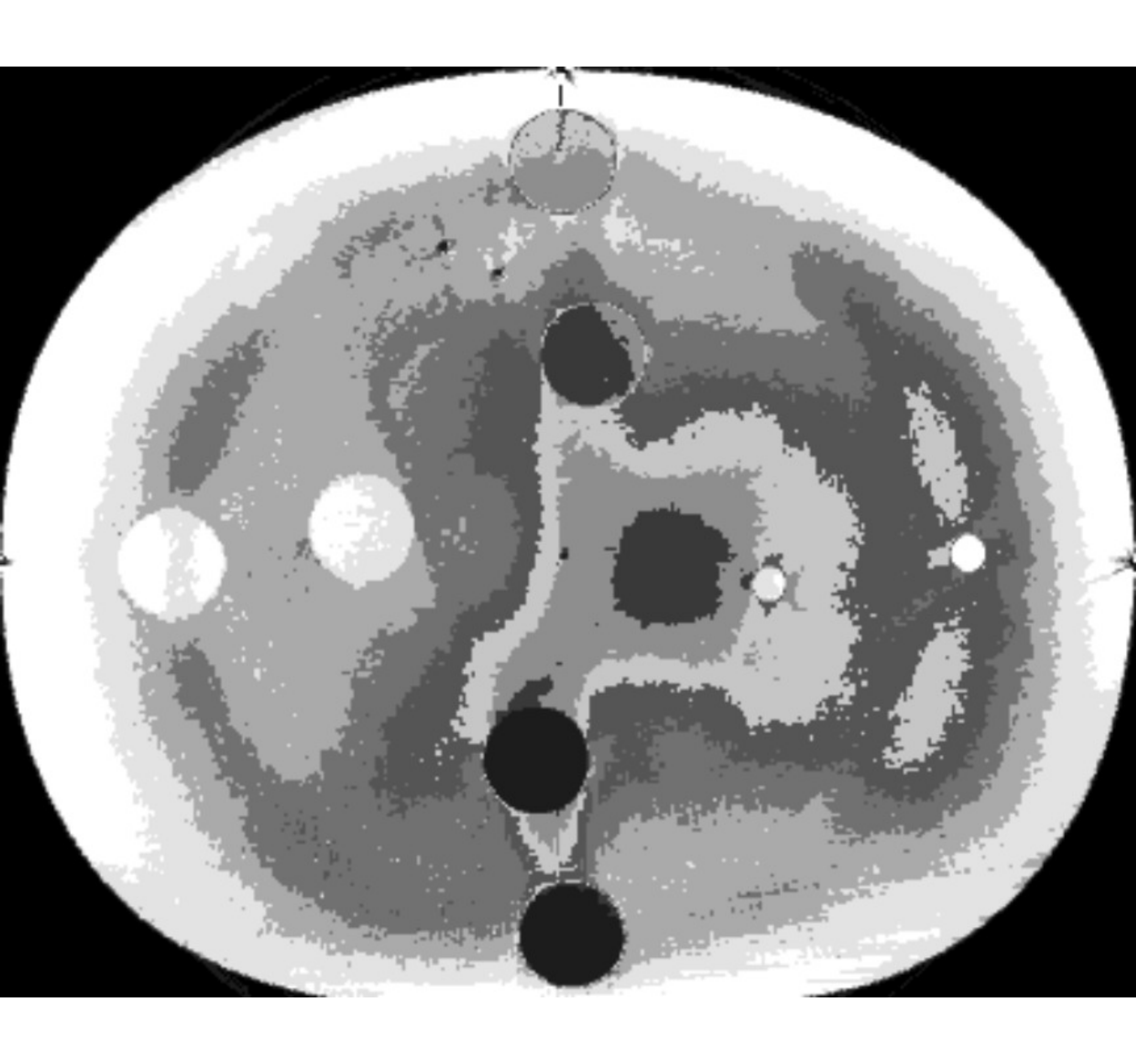}}
\qquad
  \subfloat[Segmentation (with external field prior)]{\label{f:res_seg} \includegraphics[width=57mm]{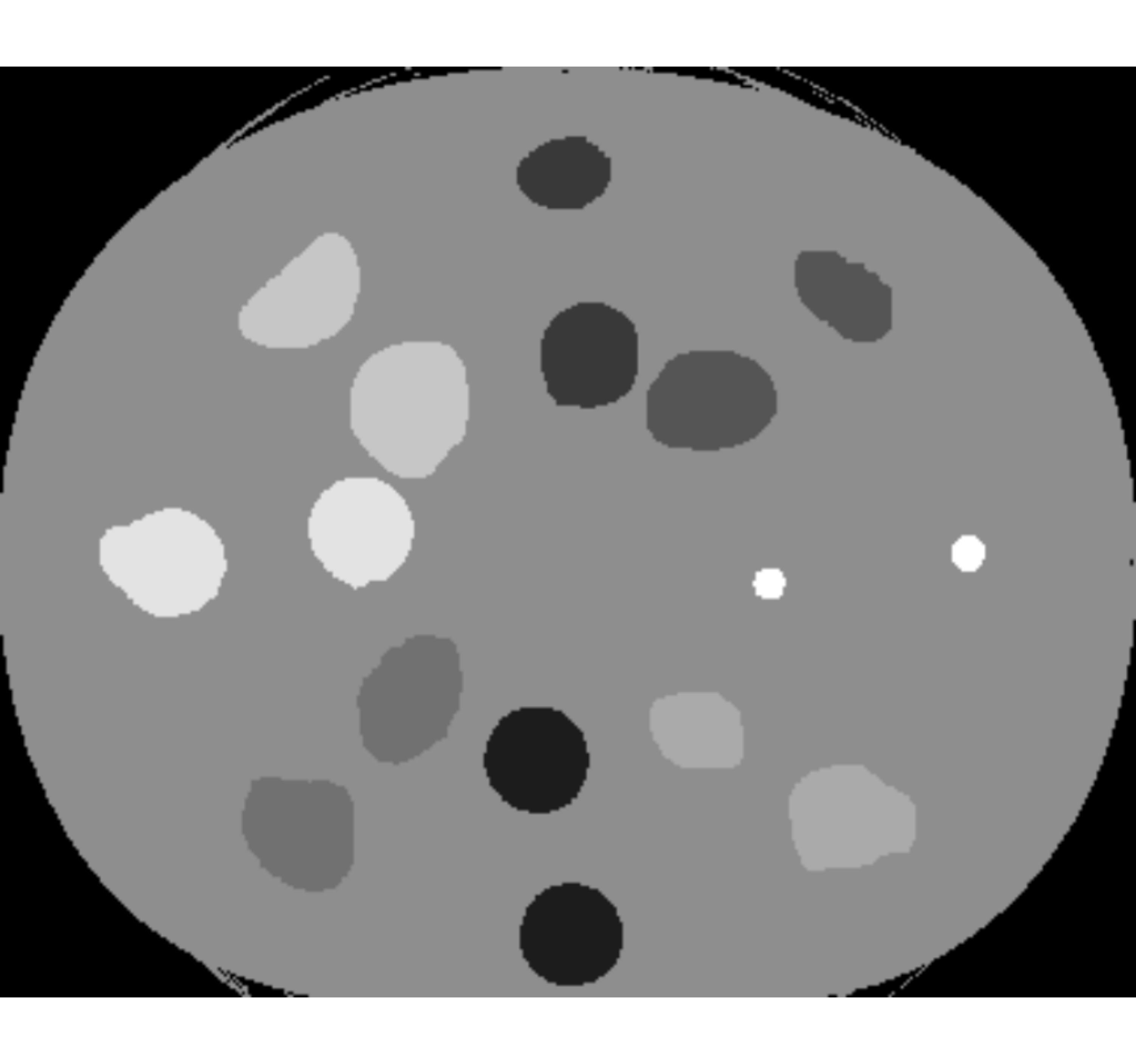}}
% figure caption is below the figure
\caption{An example segmentation result for a cone-beam CT scan of the ED phantom, both with and without the external field prior.}
\label{f:results}
\end{figure*}
We fitted the hidden Potts model to the data both with and without the external field prior, to measure the difference in the resulting segmentation accuracy. Both models were run for 55,000 MCMC iterations, discarding the first 5,000 as burn-in. This took between 9 and 14 hours per scan on a 2.66GHz Intel Xeon computer with 12GB RAM, running 64bit GNU/Linux 3.0.38 and R 2.15.3. As well as monitoring the values of the model parameters $\mu_j$, $\sigma_j^2$ and $\beta$, we also calculated the sum of like neighbours and the number of correctly-classified voxels at each iteration. These additional statistics were used as convergence diagnostics, to check that the model was mixing well and that it had reached a steady state.

The voxels were labelled according to the most frequently-assigned value of $z_i$ over the remaining 50,000 iterations. Classification accuracy was measured using the Dice similarity coefficient \citep{Dice1945}:
\begin{equation}
\label{eq:Dice}
DSC_j = \frac{2 \times | \hat\jmath \cap  j |}{| \hat\jmath | + n_j}
\end{equation}
 where
 $DSC_j$ is the Dice similarity coefficient for label $j$,
 $| \hat\jmath |$ is the count of voxels that were classified with the label $j$,
 $n_j$ is the number of voxels that are known to truly belong to component $j$, and
 $| \hat\jmath \cap j |$ is the count of voxels in $j$ that were labeled correctly.

 The results are summarised in Table~\ref{t:results}. Individual results for each cone-beam CT scan are available in the online supplementary material. An example of one of the segmentations is illustrated in Fig.~\ref{f:res_seg}. When we estimated the effect size of the external field prior using paired differences, we found that it was dependent on the tissue type. For soft tissue (adipose, breast, liver and muscle) the 95\% highest posterior density (HPD) interval for the paired differences in Dice coefficient (with and without the external field) was between 0.64 and 0.66. The smallest improvement was between 0.34 and 0.38 for lung (inhale). Overall, the mean voxel misclassification rate improved from 86.8\% to 6.2\%.

% For tables use
\begin{table}
% table caption is above the table
\caption{Segmentation Accuracy ($DSC_j \pm \sigma$)}
\label{t:results}       % Give a unique label
% For LaTeX tables use
\begin{tabular}{lrr}
\hline\noalign{\smallskip}
Tissue Type & Simple Potts & External Field  \\
\noalign{\smallskip}\hline\noalign{\smallskip}
Lung (inhale) & $0.540 \pm 0.037$ & $0.902 \pm 0.009$ \\
Lung (exhale) & $0.172 \pm 0.008$ & $0.814 \pm 0.022$ \\
Adipose       & $0.059 \pm 0.008$ & $0.704 \pm 0.062$ \\
Breast        & $0.077 \pm 0.011$ & $0.720 \pm 0.048$ \\
Water         & $0.174 \pm 0.003$ & $0.964 \pm 0.003$ \\
Muscle        & $0.035 \pm 0.004$ & $0.697 \pm 0.076$ \\
Liver         & $0.020 \pm 0.007$ & $0.654 \pm 0.033$ \\
Spongy Bone   & $0.094 \pm 0.014$ & $0.758 \pm 0.018$ \\
Dense Bone    & $0.014 \pm 0.001$ & $0.616 \pm 0.151$ \\
\noalign{\smallskip}\hline
\end{tabular}
\end{table}

The posterior estimate of the inverse temperature was almost identical for all 27 cone-beam CT scans. Without the external field prior, the pooled 95\% HPD interval for $\beta$ was $[0.793; 0.795]$. With the prior, the HPD interval for $\beta$ converged to $[1.151; 1.197]$. As discussed in Section~\ref{s:sensitivity}, the value of $\beta$ changes with the strength of the prior, as measured by the standard deviation.

\subsection{Sensitivity analysis}
\label{s:sensitivity}
\begin{figure}
   \includegraphics[width=5.3in]{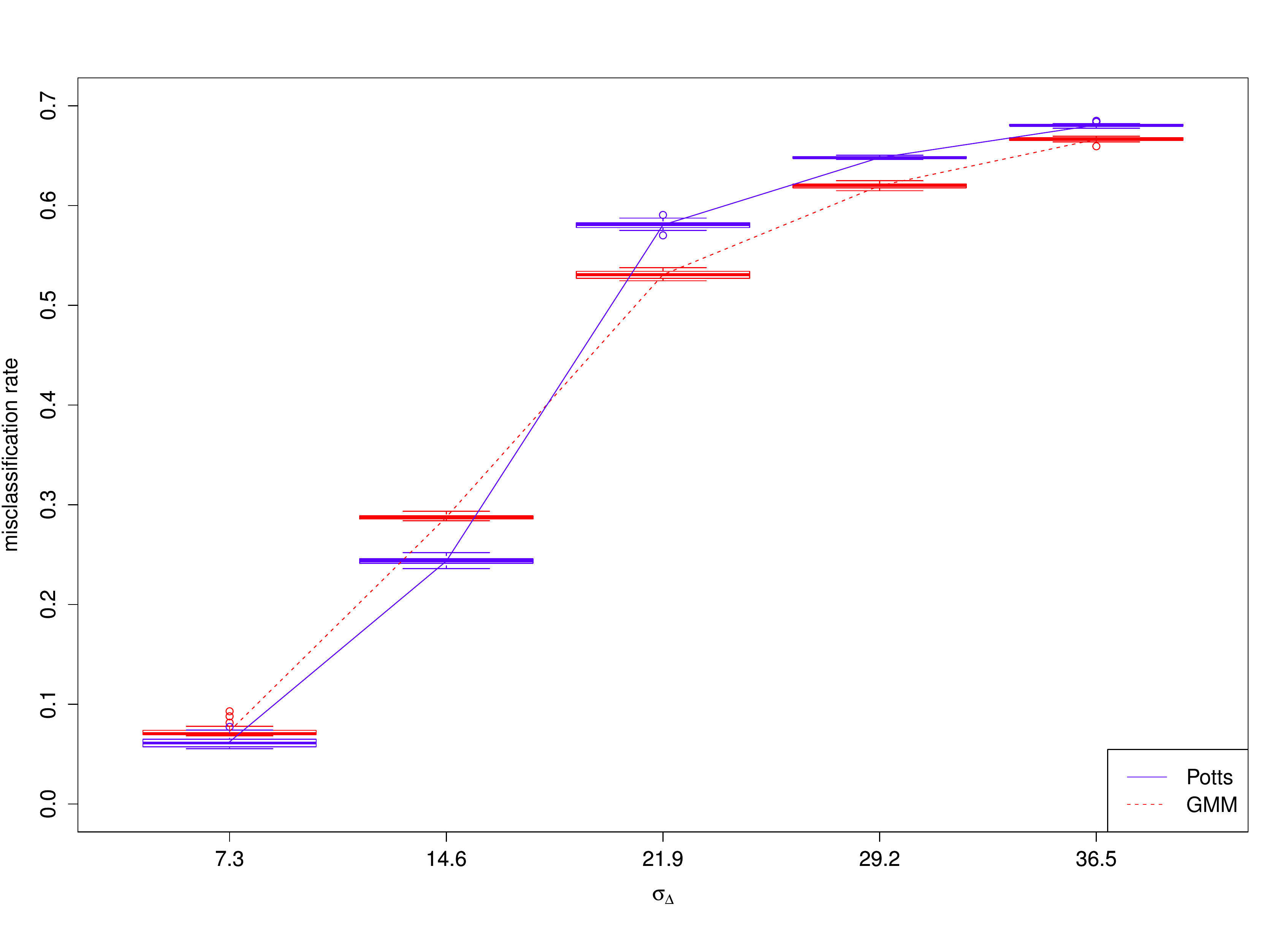}
\caption{The effect of uncertainty (standard deviation) in the external field prior on segmentation accuracy obtained from 27 cone-beam CT scans, comparing the hidden Potts model to an independent mixture of Gaussians ($\beta=0$).}
\label{f:xfield_misclass}
\end{figure}
The amount of spatial variability in our application is quite small (standard deviation of $7.3\textrm{mm}$) relative to the size of the images ($330\times270\times46\textrm{mm}$). We reran the model using external field priors with increasing levels of uncertainty ($2\sigma_\Delta$ up to $5\sigma_\Delta$, illustrated in Fig.~\ref{f:xfield_sigma}) to measure the relationship between the strength of the prior and the resulting segmentation accuracy. The pixel-wise misclassification rates for the 27 cone-beam CT scans are shown in Fig.~\ref{f:xfield_misclass}. We found that there was a strong relationship, with the proportion of misclassified pixels increasing sharply at first, but then gradually levelling off. As the uncertainty increases, we would expect that the misclassification rate would approach 86.8\%, the accuracy that is observed without the external field prior.

Fixing the value of $\beta$ at the point estimate of 1.2 did not have any significant effect on segmentation accuracy (mean pairwise difference of 0.1\% with standard deviation of 0.1). In terms of computational cost for path sampling, there was a mean pairwise difference of 9 hours (standard deviation 4.3 hours) in CPU time. However, this did not make any significant difference to the elapsed runtime, due to the variability in execution of the multi-threaded process.

We also measured the effect on segmentation accuracy when $\beta$ was fixed at zero, which is equivalent to an independent mixture of Gaussians. When $\sigma_\Delta=7.3\textrm{mm}$ the average misclassification rate was 7.3\% (mean pairwise difference of 1.07\% with standard deviation of 0.03). This indicates that the Potts model contributed very little to the result. Most of the spatial dependence in the model was accounted for by the external field prior, while the local smoothing of the labels made a small but significant difference. When $\sigma_\Delta$ was doubled to $14.6\textrm{mm}$, the segmentation accuracy decreased overall but the contribution from the Potts model was larger (mean pairwise difference of 4.41\% with standard deviation of 0.21). When the uncertainty was increased still further, the Potts model actually made the accuracy worse. This can be seen in Fig.~\ref{f:xfield_misclass} for $\sigma_\Delta \in \{ 21.9, 29.2, 36.5 \}$. This is because the Potts model relies on the labels of the neighbouring pixels. If many of the pixels are misclassified, this will then bias the probabilities of the labels towards an incorrect value.

\begin{figure}
   \includegraphics[width=5.3in]{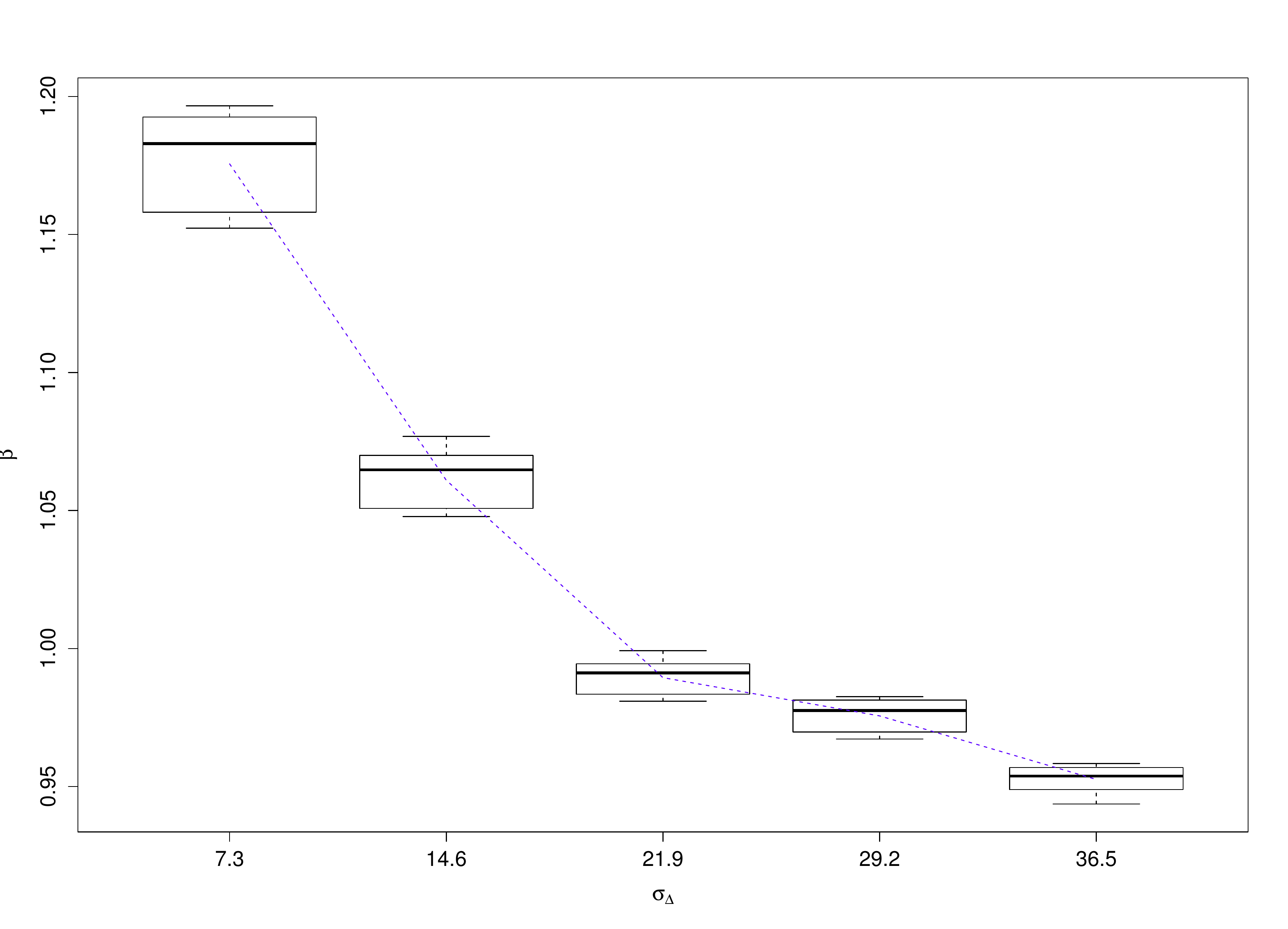}
\caption{The effect of uncertainty in the external field prior on the posterior mean of the inverse temperature for 27 cone-beam CT scans.}
\label{f:xfield_beta}
\end{figure}
The value of the inverse temperature obtained by path sampling also changed as the uncertainty in the prior increased. Fig.~\ref{f:xfield_beta} shows the posterior mean of $\beta$ for the 27 cone-beam CT scans. It is evident from Eq.~(\ref{eq:post_Potts}) that $\beta$ is a free parameter that balances the strength of spatial association against the external field (which in our model is data $\times$ prior). The external field prior becomes weaker as the standard deviation increases, thus the value of $\beta$ is lower to compensate.

\section{Discussion}
\label{s:conclusion}

We have proposed a novel formulation for introducing additional spatial prior information into the hidden Potts model, via an external field. We fit this model using a combination of Gibbs sampling, chequerboard updating and path sampling. Our implementation of these algorithms is available in the supplementary material. The method is particularly applicable in longitudinal imaging, where the segmentation of one image can be used as a prior for the next. In this paper we have used the same prior for every image, since we focused on comparing the results that were obtained when the prior was not used. In a longitudinal setting it would also be possible to update the prior dynamically as each image is processed, using a similar approach to \citet{Alston2007} for the mean and variance.

In situations where the intensity values and spatial homogeneity are insufficient for accurate classification of the image pixels, an external field prior substantially improves segmentation accuracy. We demonstrated our method by applying it to cone-beam CT scans, which are medical images that are used for image-guided interventions, such as external-beam radiotherapy. The external field prior improved the pixel misclassification rate from 87\% to 6\%.

This method shows great potential for application in automated analysis of daily cone-beam CT scans of radiotherapy patients. By deriving the external field prior from a patient's treatment plan, the method would better accommodate within-patient variation than the probabilistic anatomical atlases currently in use. The method could also be applied to other images, such as MRI or satellite remote sensing. These applications involve repeated observations of a subject over time. In order for the external field prior to be most effective, the rate of change in the subject should be gradual, relative to the frequency with which the process is observed.

Due to the average runtime of ten hours, the MCMC algorithms described in this paper would be unsuitable for online use during patient treatment. However, they could be used offline for dose tracking and other monitoring activities. Approximate algorithms such as iterated conditional modes \citep[ICM;][]{Besag1986} or variational Bayes \citep[VB;][]{McGrory2009} could also be used to fit the model in time-critical applications such as image-guided radiotherapy. In \citet{Moores2014} we have previously shown that the hidden Potts model with external field prior can be fit using ICM with an average runtime of only 9 minutes. A drawback of these methods is that the options for estimating the inverse temperature are more limited, although a fixed value of $\beta$ would be sufficient for most purposes.

\section*{Acknowledgements}
The authors thank Louise Ryan, Nial Friel and Anthony Pettitt for useful discussions, as well as the anonymous reviewers for their comments on earlier versions of this work. We would also like to thank the staff of the Radiation Oncology Mater Centre, particularly Emmanuel Baveas, for their assistance with this project. Moores thanks Queensland University of Technology and the Australian federal government Department of Education, Science and Training for financial support. Mengersen's research is funded by a Discovery Project grant from the Australian Research Council. Computational resources and services used in this work were provided by the HPC and Research Support Group, Queensland University of Technology, Brisbane, Australia.

\bibliographystyle{elsarticle-harv}
\bibliography{xfield_prior}

\end{document}